\documentclass[aps,reprint,amsmath,amssymb]{revtex4-2}

\usepackage[T1]{fontenc}
\usepackage{graphicx}
\usepackage{amsmath}
\usepackage{gensymb}
\usepackage{upgreek}
\usepackage{xcolor}
\usepackage{tabularx}
\usepackage{array}

\begin{document}

\title{Interacting Random-field Dipole Defect Model for Heating in Semiconductor-based Qubit Devices}
\author{Yujun Choi}
\affiliation{Department of Physics, University of Wisconsin-Madison, Madison, WI 53706, United States}

\author{Robert Joynt}\email{rjjoynt@wisc.edu}
\affiliation{Department of Physics, University of Wisconsin-Madison, Madison, WI 53706, United States}

\date{\today} 

\begin{abstract}
Semiconductor qubit devices suffer from the drift of important device parameters as they are operated.  The most important example is a shift in qubit operating frequencies.  This effect appears to be directly related to the heating of the system as gate operations are applied.  We show that the main features of this phenomenon can be explained by the two-level systems that can also produce charge noise, if these systems are considered to form an interacting random-field glass.  The most striking feature of the theory is that the frequency shift can be non-monotonic in temperature.  The success of the theory narrows considerably the possible models for the two-level systems.
\end{abstract}

\maketitle

\section{Introduction}
There has been considerable progress in semiconductor quantum computing, with significant strides in scaling up and in gate fidelities \cite{philips2022universal,mills2022two}. The chief difficulty is decoherence, with charge noise as the main culprit \citep{DekkerPRL1991, sakamoto1995distributions, KurdakPRB1997, Buizert:2008p1511, PioroPRB2005, Takeda:2013p123113, croot2017gate, campbell2002density, hofheinz2006individual}. The precise nature of the two-level systems (TLS) that produce the noise remains elusive. There have been extensive characterizations of the noise spectrum  \cite{connors2019low, connors2022charge}, and the spatial correlations in the noise \cite{boter2020spatial, Rojas-Arias} in different devices.  These observations can help in the elucidation of the nature of the TLS, but the data are not currently sufficient to pin things down precisely.

Another problem, at first sight quite separate, that interferes with qubit operation is the pulse-induced resonance shift (PIRS).  This is a shift in the operating frequency of the qubits as a computation proceeds.  This is highly problematic, since continual recalibration of the system is not practical.  Quadrature control \cite{takeda2018optimized} and prepulsing \cite{Watson:2018p633} can mitigate but not eliminate this issue.  PIRS has recently received an intensive experimental study \cite{Undseth}.  

Here we propose that the source of PIRS is also a group of TLS, perhaps the same group that gives charge noise.  Hence detailed observations of PIRS may provide additional insight into the microscopic origin of the noise.  We proceed in the time-honored fashion of proposing a phenomenological model that explains the data, and then seeing what constraints the model places on the underlying physics of the system.

Let us take a concrete situation in which the system is in a resting state at low temperature $T$ for times $t<0$ and the operations, which involve microwave pulses that feed energy into the system, begin at $t=0$ and end at some later time $t_f$. PIRS is a time-dependent shift $\Delta f(t)$ with $\Delta f(t=0) = 0$ by definition. $\Delta f$ is a function of time that ultimately reverts to the base state some time after the operations have ceased. PIRS appears to be rather ubiquitous in semiconductor qubits, but there is considerable variability in how it manifests itself. Early observations found positive shifts ($\Delta f \geq 0) $ of order a few MHz \cite{Watson:2018p633}. The magnitude of the shift was an increasing function of the energy injected by the pulses. It also depended on the details of the electron wavefunctions in the dots, for example on the dot occupations.  The MHz magnitude of the shifts is fairly typical for quantum dot qubits.  Importantly, $\Delta f$ can also be negative \cite{takeda2018optimized,zwerver2022qubits}.  The decay time after $t_f$ varies in the dot systems, with values from 0.5 ms \cite{philips2022universal} to 38 $\mu$s \cite{zwerver2022qubits} having been observed.  PIRS also occurs in donor-based qubit systems \cite{freer2017single}, though $|\Delta f|$ is much smaller, of order 10s of kHz.  Effects of a similar magnitude are seen in flip-flop qubits \cite{savytskyy2022electrically}.  In this work we concentrate on experiments in dots, but we expect the theory to apply more broadly.

Our interpretation is based on the fact that dot and donor systems share the feature that the qubit operating frequency depends on the spatial position of the qubit. The arbitrary sign of $\Delta f$ then suggests that a change in the electric field on the qubit is the origin of PIRS. Experimentally, it now appears to be clear that PIRS is essentially a thermal heating effect rather than a mechanical effect \cite{philips2022universal, Undseth}.  This is also supported by the characteristic return to a base state, most naturally interpreted as a return to thermal equilibrium. The most striking feature of the results is that the magnitude of $\Delta f$ is typically not monotonic in temperature ($T$), instead rising to a maximum at about 200-300 mK, then decreasing \cite{Undseth}.

\section{Model}
To explain these observations, we introduce a model based on the charged TLS that are known to exist in these devices, and that in fact are also responsible, at least in part, for the decoherence of the qubits.  These charged defects or traps are modeled as a collection of $N$ fluctuating electric dipoles.  The $j$th dipole fluctuates between states $s_j \mathbf{p}_j$, where $s_j = \pm 1$ and $\mathbf{p}_j$ is a fixed vector for each $j$. For simplicity we assume that the dipoles all have the same magnitude: $|\mathbf{p}_j| = p_0$.  This is reasonable if all the TLS have the same physical origin.  The dipoles can have a non-zero equilibrium moment which is random in direction and they interact via the long-range Coulomb interaction.  We call this the Interacting Random-field Glass Model (IRGM). Somewhat similar models have been introduced to understand charge noise (rather than equilibrium fields) in dot systems \cite{mickelsen2023effects} and also in the context of superconducting qubit systems to explain fluctuations in the relaxation time $T_1$ \cite{muller2015interacting,muller2019towards}.

The electric field $\langle \mathbf{F}_q \rangle$ on a qubit at the origin of coordinates is
\begin{equation}
\label{eq:Fq}
\langle \mathbf{F}_q \rangle =
\frac{1}{4 \pi \varepsilon}
\sum_{j=1}^{N} \langle s_j \rangle 
\frac{3 (\mathbf{p}_j \cdot \mathbf{r}_j) \, \mathbf{r}_j - \mathbf{p}_j |\mathbf{r}_j|^2}
{ |\mathbf{r}_j|^5}
\equiv \sum_{j=1}^{N} \langle s_j \rangle \mathbf{F}_j.
\end{equation}
The angle brackets indicate a thermal average.  In the devices in question, the qubit operation frequency depends linearly on the electric field at its position. The frequency is a quasi-equilibrium quantity, so  $\langle \mathbf{F}_q \rangle $ is the object of interest for our purposes.  The relation between field and frequency is platform-dependent.  In the set-up of Refs.~\cite{Watson:2018p633, takeda2018optimized, philips2022universal}, $\langle \mathbf{F}_q \rangle$ causes the displacement of the spin qubit in a magnetic field gradient, while in the flip-flop qubit architecture motion of the qubit caused by  $\langle \mathbf{F}_q \rangle$ would change the hyperfine coupling or the g-factor \cite{Rahman_PRB2009}.  In all cases the displacement changes the qubit operating frequency. The qubit frequency is
$f(T) = f_0 + \mathbf{c}_q \cdot \langle \mathbf{F}_q \rangle (T=0) + \Delta f(T)$, where $f_0$ is the $T$-independent part from the applied magnetic field, $\langle \mathbf{F}_q \rangle (T=0) \neq 0$ is a constant that comes from the ground state configuration of the TLS, and $\Delta f(T)$ is the PIRS effect and all the $T$ dependence of $f$ comes from it.
Thus $\Delta f(T) = \mathbf{c}_q \cdot [ \langle \mathbf{F}_q \rangle (T) - \langle \mathbf{F}_q \rangle (T=0)]$ is the quantity of interest. $\mathbf{c}_q$ depends on the particular type of qubit and the position of the qubit in the device. Its direction is determined by the condition that the effective magnetic field produced by $\langle \mathbf{F}_q \rangle$ 
should be parallel to the applied field.

The Hamiltonian of the TLS in our model contains a random-field term $H_r$ and an interaction term $H_{int}$: 

\begin{equation*}
   H = H_r + H_{int}=-p_0 \sum_{j=1}^N s_j \mathbf{E}_j \cdot \hat{p}_j - \frac{p_0^2}{8 \pi \varepsilon} \sum_{j \neq k=1}^N s_j s_k V_{jk}. 
\end{equation*}
Here
\begin{equation*}
\label{eq:Vjk}
    V_{jk} = \frac{3 \hat{p}_j \cdot (\mathbf{r}_j - \mathbf{r}_k) \,
    \hat{p}_k \cdot (\mathbf{r}_j - \mathbf{r}_k)
    - (\hat{p}_j \cdot \hat{p}_k) \, |\mathbf{r}_j - \mathbf{r}_k|^2
    }
    {|\mathbf{r}_j - \mathbf{r}_k|^5}.
\end{equation*}

The random effective electric fields $\mathbf{E}_j$ if interpreted in a double-well picture of the TLS are related to the energy asymmetry (`detuning') of the two wells.  However, the physical origin of the $\mathbf{E}_j$ may not be the same in all cases.  For example, they could be actual external electric fields coming from the gate electrodes, strain fields, asymmetric microscopic defects, \textit{etc}.  For our purposes they are considered to be phenomenological parameters that must be fit, since they are very difficult to estimate in the absence of a real microscopic model.
We expect $N$ to be a number in the range of perhaps 10 to 100 and to be sample-dependent \cite{connors2019low}.
The dipoles may well be the same TLS that give rise to the noise in the system, but here we are interested in their equilibrium behavior, not their fluctuations.  This assumes that measurement of PIRS takes place over a time interval longer than the characteristic switching times of the TLS.  However, the intersection of the set of TLS that causes qubit dephasing and the set that causes PIRS need not be complete.

\section{Non-monotonicity of PIRS}

The $T$ dependence of $\langle \mathbf{F}_q \rangle$ in a non-interacting model with $H_{int} = 0$ is already interesting, so we discuss it in detail.  In this case the problem is exactly solvable, once the positions of the dipoles are specified: $ \langle s_j \rangle = \textrm{sgn}(\hat{p}_j \cdot \mathbf{E}_j) \tanh(p_0 \, \hat{p}_j \cdot \mathbf{E}_j / k_B T)$.  $\langle s_j \rangle$ has a definite sign at $T=0$ but eventually $\langle s_j \rangle \rightarrow 0$ as $T \rightarrow \infty$. We can identify a turn-off temperature $T_j= p_0 |\mathbf{E}_j| / k_B$ for each TLS.  Substitution into Eq.~\ref{eq:Fq} and performing the sum gives the equilibrium electric field $ \langle \mathbf{F}_q \rangle (T)$ at the qubit.

To understand the qualitative $T$ dependence of $ \langle \mathbf{F}_q \rangle$ in the IRGM  we begin with $T=0$. We divide the TLS into two groups.  In the set $S^+$ we have the indices $j$ for which  $  \langle \mathbf{F}_q \rangle (T=0)\cdot \mathbf{F}_j 
\langle s_j \rangle (T=0) >0$ while in group $S^-$ we have the indices $j$ for which
$  \langle \mathbf{F}_q \rangle (T=0) \cdot \mathbf{F}_j \langle s_j \rangle (T=0) < 0$.  That is, the dipoles in $S^+$ are aligned with the ground state resultant field, while those in $S^-$ are anti-aligned.  The electric field at the qubit is the result of a random walk of the vectors $\langle s_j \rangle \, \mathbf{F}_j$ with a resultant vector 
\begin{equation}
    \langle \mathbf{F}_q \rangle 
    = \sum_{j\in S^+} \langle s_j \rangle \mathbf{F}_j 
    + \sum_{j\in S^-} \langle s_j \rangle \mathbf{F}_j 
\end{equation}
The vectors in group $S^+$ are in the direction of the final result of the walk, while those in group $S^-$ are in the opposite direction.  Due to the randomness in the asymmetry, the various components of the walk turn off at different temperatures, and there will be some average turn-off temperature $T^+$ for group $S^+$ and a different average turn-off temperature $T^-$ for group $S^-$.  We now increase $T$ from zero. If $T^+ > T^-$  then the total field strength $ | \langle \mathbf{F}_q \rangle | $ will first increase and eventually vanish when $T \gg T^+$.  In this case we have a non-monotonic $T$-dependence of the field strength $ | \langle \mathbf{F}_q \rangle | $, a rather surprising result. 
  $T^+$ and $T^-$ are not expected to be very different if the dipoles have the same physical structure, but even in this case the relatively small value of $N$ implies that random fluctuations will make $T^+ \neq T^-$.  If $T^+ < T^-$  then the total field strength $ | \langle \mathbf{F}_q \rangle | $ will decrease as the dominant dipoles turn off and eventually vanish when $T \gg T^-$.  If there is a gross mismatch between the  $T^+$ and $T^-$ components of $ | \langle \mathbf{F}_q \rangle | $ could even reverse sign, but overall we would expect a monotonic decrease.  The relative magnitudes of $ |\sum_{j\in S^+} \langle s_j \rangle \mathbf{F}_j |$ and $| \sum_{j\in S^-} \langle s_j \rangle \mathbf{F}_j|$ are also important.  If $ |\sum_{j\in S^+} \langle s_j \rangle \mathbf{F}_j |$ dominates, then there is little cancellation in the sum and the non-monotonic behavior will be suppressed.  If the fields from $S^+$ and $S^-$ are comparable, then non-monotonicity is more likely.   

Now we turn to the effects of interactions. Overall, dipolar interactions are always antiferromagnetic.  This favors depolarization - a smaller net moment $ | \langle \mathbf{P} \rangle | = | \sum_j \langle s_j \rangle  \mathbf{p}_j | $.  If the TLS are located to one side of the qubit, then the correlation between $ | \langle \mathbf{P} \rangle |$ and $ | \langle \mathbf{F}_q \rangle |$ will be strong, but even if the qubit is surrounded symmetrically by the TLS, fluctuations will still give some correlation in a given sample.  Small $ | \langle \mathbf{P} \rangle | $ comes from cancellation in the directions of the individual moments. 

There is an additional temperature scale associated with the interactions, which is the average change in the interaction energy from flipping one spin: $ T_{int} = |\langle H_{int} \rangle | / N k_B$.  If $T_{int} < T^+,T^-$ then the interactions will turn off before the random field effects as $T$ is increased, and $ | \langle \mathbf{P} \rangle | $ increases. If $T_{int} \gg T^+,T^-$ then the system is frozen by the interactions and we expect little change in $ | \langle \mathbf{P} \rangle | $ until $T \gg T_{int}$.

Overall, the effect of interactions is to make cancellation more likely due to their antiferromagnetic character.  Unless the interactions are extremely strong, they make the non-monotonic $T$ dependence of $ | \langle \mathbf{F}_q \rangle | $ and therefore of $\Delta f(T) $ more likely.  

\section{Numerical Results}
This analytic analysis of the IRGM is semi-quantitative.  To make the arguments more firm we perform numerical simulations.  We do this for three different physical pictures of the TLS. In all simulations there is a single qubit at the origin.

In the first picture the TLS are charge traps near the surface of a two-dimensional electron gas.  The trap is  positively charged when empty and then relatively negatively charged when full.  We include the image charge.  This can be described as a fluctuating dipole in the z direction, z being the growth direction.  These dipoles are uniformly distributed in a thin layer at positions $\mathbf{r}_j = (x_j,y_j,z_j)$ with $-150 \, \mathrm{nm} < x_j, y_j < 150 \, \mathrm{nm}$ and $z_j= 50 \, \mathrm{nm}$.  This is the trap picture.

The second picture conceives of the TLS as point defect dipoles in the oxide with orientations uniformly distributed in direction.  Their positions  $\mathbf{r}_j = (x_j,y_j,z_j) $ are uniformly distributed in a layer with coordinates satisfying  $\mathbf{r}_j = (x_j,y_j,z_j)$ with $-150 \, \mathrm{nm} < x_j, y_j < 150 \, \mathrm{nm}$ and $30 \, \mathrm{nm} < z_j < 50 \, \mathrm{nm}$.  This is the random dipole picture. 

In the third picture the TLS are distributed in the neighborhood of the qubit.  For the simulations, we take the TLS to be uniformly distributed in a spherical shell with the qubit at the center. $\mathbf{r}_j = (r_j, \theta_j, \phi_j) $ satisfy  $60 \, \mathrm{nm} < r_j < 80 \, \mathrm{nm}$,  $0 \leq \theta_j < \pi$, and $0 \leq \phi_j < 2\pi$.  The radii are chosen to make $T_{int} \sim 1$ K.  This is the spherical shell picture.

\begin{figure*}[t]
\centering
\includegraphics[width=\textwidth]{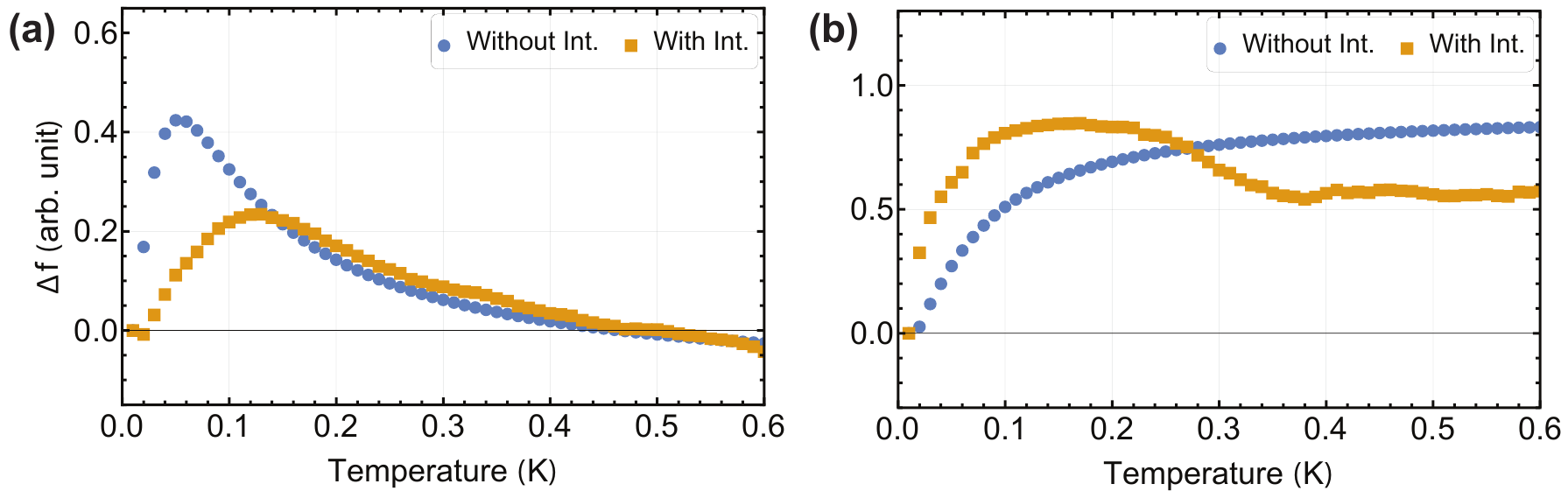}
\caption{Temperature dependence of qubit frequency shifts in the trap picture of the TLS in some representative samples.  The shifts are proportional to one component of the equilibrium electric field $\langle \mathbf{F}_q \rangle (T)$ at the qubit.  Without Int. and With Int. mean the non-interacting and fully interacting case, respectively.  \textbf{(a)} Both non-interacting and interacting cases show non-monotonic shifts. \textbf{(b)} Example in which the interaction causes non-monotonic behavior.}
     \label{fig:trap}
\end{figure*}

We define a random field temperature scale $ T_{r} = |\langle H_{r} \rangle | / N k_B$ in addition to $T_{int}$. We use the parameter values $p_0 = 48$ Debye $\approx 1 |e|$-nm and $N=30$, which are chosen because they give $T_2$ of order $10^{-6}$s in the correct experimental range \cite{choi2022anisotropy}. The TLS density from these values is also consistent with those computed from the magnitude of measured power spectra \cite{shehata2023modeling, kkepa2023correlations}. The random field strength is taken as $\Delta E_0 = 10^4$ V/m, which denotes the standard deviation of a distribution centered on zero.  With these values, $T_r \sim 0.1$ K and  $T_{int} \sim 1$ K in most samples, but while $T_r$ is almost independent of the disorder because its strength is roughly fixed, $T_{int}$ can vary from 0.1 K to 10 K because it depends sensitively on the separations and orientations of the TLS.  Overall, these values are chosen to be representative of experiments on semiconductor qubit systems, in the sense that most analyses give something like our value for $N$ which together with value of $p_0$ that we use gives a reasonable dephasing time. A key observation here is that $T_{int}$ comes out approximately correct when the distances of the TLS from each and from the qubit, and the dipole magnitude, are chosen to fit $T_2$ and other noise experiments. Some further details are given in Appendix \ref{magnitude}.
The IRGM passes the self-consistency check that $T_{int}$ does indeed match the temperature scale on which $\Delta f $ varies. 

In Fig.~\ref{fig:trap} we plot one example of $ \Delta f (T)$ in the trap picture for both the non-interacting case $H_{int}=0$, which is exactly solvable, and the fully interacting case, which is computed by a standard Monte Carlo (MC) algorithm.  We use arbitrary units for $\Delta f$ since the conversion factor $\mathbf{c}_q$ is platform-dependent.  For the interacting case, a moving average over 11 neighboring temperatures is applied to obtain stable results, and a smaller number of neighboring temperatures are used for the moving average at the ends of the curves.

We stress that $ \langle \mathbf{F}_q \rangle (T) $ is sample-dependent for all three pictures in that changing the parameters in natural ranges can alter $\Delta f(T)$ qualitatively.  In particular, non-monotonic behavior of $\langle \mathbf{F}_q \rangle $ is by no means universal.  It is even possible for a single sample that one component of the field is non-monotonic and another is monotonic.  Some idea of the variety of possible behaviors is 
given in Appendix \ref{examples}.

Fig.~\ref{fig:trap} demonstrates that non-monotonicity can arise already even when the TLS do not interact.  This comes simply from the fact that there can be cancellation of the random fields at $T=0$ that is lessened as $T$ increases in certain circumstances, as explained above.  The interactions tend to enhance non-monotonicity as in Fig.~\ref{fig:trap}(a), though this effect is not universal. In fact, as we will see below, only a minority of samples for the trap picture show non-monotonicity.   
Interactions can create non-monotonicity when the non-interacting picture shows monotonicity as seen in Fig.~\ref{fig:trap}(b).  Again, this is consistent with the idea of cancellation as the active ingredient in non-monotonicity.  Additional examples of the different types of behavior that can occur for $\Delta f(T)$ are given in Appendix \ref{examples}.      

Once universality of the non-monotonicity is ruled out, the question becomes whether it is likely or not.  To answer this we did simulations over 10,000 samples for each of the three pictures and determined whether a monotonic or non-monotonic behavior was observed for each component of the electric field. The precise criterion for monotonicity or its absence is given in Appendix \ref{nonmono}.  

\begin{table}[t]
\centering
\setlength\doublerulesep{0.1in} 
\begin{tabular}{| c | c | c | c |}
 \hline
 Trap & $\langle F_{q,x} \rangle$ & $\langle F_{q,y} \rangle$ & $\langle F_{q,z} \rangle$ \\
 \hline
 \multicolumn{4}{|c|}{$T_r < T_{int}$} \\
 \hline
  Without Int. & 15.9\% & 16.4\% & 12.0\% \\
  With Int. & 48.0\% & 48.6\% & 40.1\% \\
 \hline
 \multicolumn{4}{|c|}{$T_r \sim T_{int}$} \\ 
 \hline
  Without Int. & 11.0\% & 10.9\% & 7.4\% \\
  With Int. & 30.5\% & 30.2\% & 24.2\% \\

\hline
\hline

 Random dipole & $\langle F_{q,x} \rangle$ & $\langle F_{q,y} \rangle$ & $\langle F_{q,z} \rangle$ \\
 \hline
 \multicolumn{4}{|c|}{$T_r < T_{int}$} \\
 \hline
  Without Int. & 14.7\% & 15.9\% & 12.4\% \\
  With Int. & 61.9\% & 61.9\% & 57.2\% \\
 \hline
 \multicolumn{4}{|c|}{$T_r \sim T_{int}$} \\ 
 \hline
  Without Int. & 10.0\% & 9.4\% & 7.8\% \\
  With Int. & 42.5\% & 42.3\% & 36.9\% \\

\hline
\hline

 Spherical shell & $\langle F_{q,x} \rangle$ & $\langle F_{q,y} \rangle$ & $\langle F_{q,z} \rangle$ \\
 \hline
 \multicolumn{4}{|c|}{$T_r < T_{int}$} \\
 \hline
  Without Int. & 25.0\% & 25.3\% & 28.0\% \\
  With Int. & 89.5\% & 89.4\% & 90.1\% \\
 \hline
 \multicolumn{4}{|c|}{$T_r \sim T_{int}$} \\ 
 \hline
  Without Int. & 21.0\% & 21.8\% & 25.8\% \\
  With Int. & 79.4\% & 79.0\% & 81.4\% \\

\hline
\end{tabular}

\caption{The fractional number of samples showing non-monotonicity obtained from Monte Carlo simulation and exact solutions for three physical pictures. $\langle F_{q,i} \rangle$ is the $i$ component of the vector $\langle \mathbf{F}_q \rangle (T)$.  Without Int. and With Int. mean the non-interacting and fully interacting case, respectively.  $T_r < T_{int}$ indicates $T_r \sim$ 0.1 K and $T_{int} \sim$ 1 K, while $ T_r \sim T_{int}$ indicates $T_r \sim$ 0.1 K and $T_{int} \sim$ 0.1K.}  
\label{tab:fraction}
\end{table}

The results in Table~\ref{tab:fraction} support the physical picture explained above.  In the trap picture, the cancellation of fields from different dipoles is relatively small, since the vectors leading from the qubit to the TLS, while not collinear, generally do not make large angles with each other.  Similar statements apply to the random dipole picture, but there is some additional cancellation due to the different orientations. The most interesting result is for the spherical shell picture.  Here we see the appearance of non-monotonic behavior for non-interacting TLS in about one quarter of the cases, as would be suggested by the above arguments. With dipoles on all sides of the qubit, the cancellation effect is quite strong. 

 Interactions do promote non-monotonicity in all cases, as expected, especially when the interactions are strong:  $T_{int} > T_r$.  Overall, non-monotonicity increases as we proceed from the trap to the random dipole to the spherical shell pictures.  The interaction enhancement reinforces this pattern of non-monotonicity.

  \begin{figure}[t]
     \centering
     \includegraphics[width=0.48\textwidth]{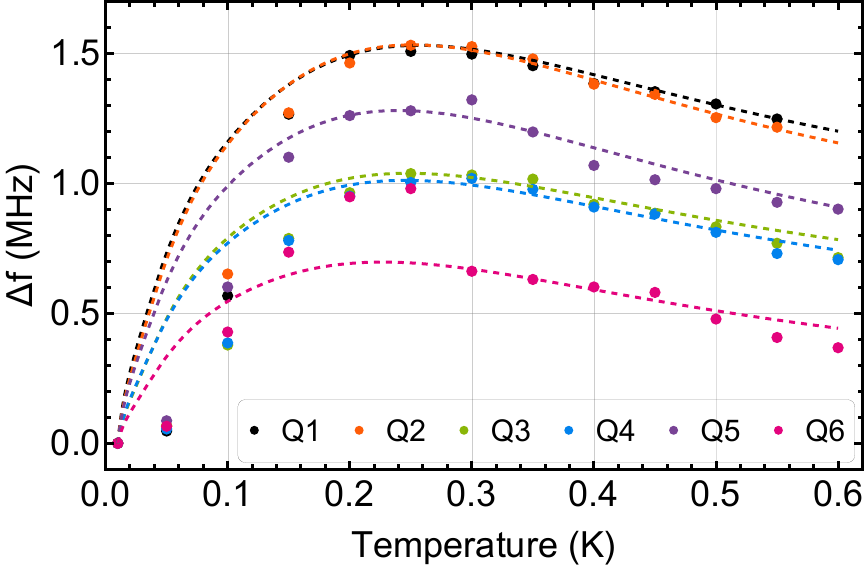}
     \caption{PIRS data: theory and experiment.  The points are measured frequency shifts for six qubits Q1-Q6 from Ref.~\cite{Undseth} and the dashed 
 lines are theoretical fits. 
 The qubits are situated in a one-dimensional array.
The fitting procedure is described in detail in the text.  The applied magnetic field is assumed to be in the $y$-direction.}
     \label{fig:fitting}
 \end{figure}

 We turn now to comparison of theory and experiment.  In nearly all measurements $\Delta f$ is measured as a function of time, not temperature.  Comparing to data of this type would require detailed modeling of the heat flow in the system, which is outside the scope of this paper.  We therefore compare only to the temperature data for $\Delta f(T)$  for 6 qubits in a single device reported in Ref.~\cite{Undseth}.  We  plot these data with a theoretical fit in Fig.~\ref{fig:fitting}.
The fit was done as follows. We first chose a set of TLS positions and random fields so that the parameters were in a range where non-monotonicity could be expected, and peak in $\Delta f$ would be around 250 mK. Then, since the curves have a fairly similar shape but differ in vertical scale, we varied $\mathbf{c}_q$ for each of the curves.  Finally, the positions of the TLS were adjusted to fit each curve individually.  The main feature that needed to be accounted for in this final step was the sharper peak and steeper falloff at high $T$ that is seen in qubits 3-6.  Some further details are given in Appendix \ref{fitting}.

The fits are quite good quantitatively, but not too much should be made of this, since the number of parameters far exceeds the number of qualitative features to be fit for each curve.  However, even given this, the fit does provide evidence for the correctness of the IRGM.  The non-monotonicity arises naturally, but also the linear behavior at small $T$, (which is caused by the uniform distribution of the $\mathbf{E}_j$ near zero) and finally the $1/T$ behavior at large $T$.  
\section{Conclusion}
We conclude by summarizing the strong and weak points of the IRGM as a model for PIRS.  There are three evaluation categories; qualitative phenomenological understanding, semi-quantitative self-consistency, and quantitative fit of theory and experiment.   

In qualitative terms, the surprising non-monotonic $T$ dependence and some of the other features of the $T$-dependence of the qubit frequencies finds a natural explanation in the IRGM.  The important theoretical ingredients are the cancellations due to vector summations that involve only a relatively small number of variables combined with the natural temperature dependence
of the TLS fluctuations and the depolarizating effects of interactions.  The explanation of non-monotonicity is quite defies the usual expectation that thermal effects in the absence of phase transitions tend to be monotonic. Overall, this is quite strong evidence for the IRGM.  Aside from the non-monotonicity, there is the observation that $\Delta f$ is sometimes negative \cite{takeda2018optimized,zwerver2022qubits}.  This is also somewhat surprising if one assumes that the heating affects the qubits directly in some fashion.  In the IRGM, there is nothing that constrains the sign of the components of $\mathbf{c}_q$, so the sign of the effect is not determined.  Similarly, the fact that the effect is not resonant with qubit frequencies suggests that an ancillary part of the device is driven by the heating - in the IRGM, the system of TLS is driven.  Hysteresis does not seem to be a feature of PIRS.  This might seem to argue against the IRGM but in fact with only a few tens of TLS involved, this aspect of glassiness does not argue against the model.  In contrast, no T-dependent electric field shows up in measurements at the charge sensor \cite{Undseth}, which is not explained in the model as it stands.   

There are two experimental scales that must be consistent with the theory: the overall magnitude of $\Delta f$ (1 MHz in dot systems), and the temperature of the peak in $\Delta f$ (about 0.2 to 0.4 K in dot systems).  The first number is very consistent with the roughly known numbers for the size of the dipoles and their presumed positions.  The second depends on the distribution of the random local fields $\mathbf{E}_j$ and indeed the distribution must be such that $| p_0 \mathbf{E}_j / k_B |$ is clustered near 0.3 K.  There is no obvious reason why this should be so, so the IRGM does include at least one \textit{ad hoc} element.

The fit to the data in Fig.~\ref{fig:fitting} is strikingly accurate, but it raises questions.  If interactions are relatively weak for some reason (such that the TLS are particularly far apart), then why do six out of six qubits all show non-monotonicity?  This is only consistent with the spherical shell picture, which in turn is not very consistent with the usual idea that the TLS are associated with the oxide layer. Furthermore, the non-monotonicity in all six qubits means that there must be correlations between the positions of the TLS and the strength and direction of their random fields.  Specifically, the TLS closer to the qubits must have stronger random fields for each qubit.  In addition, in order for the shift to be positive in all qubits, the random fields for the nearby TLS must all have the same orientation across all six qubits.      

We conclude that the basic mechanism of PIRS is explained by the IRGM, but that the explanation is far from complete at this stage.  Most likely the model needs to supplemented by a better picture of the positions of the TLS and a better understanding of their physical nature.  This would limit the model to a smaller region of its parameter space than we have investigated here.

\begin{acknowledgements}
We acknowledge helpful discussions and correspondence with S.N. Coppersmith,  M.A. Eriksson, M. Friesen, A. Laucht, A. Morello, A. Saraiva, L.M.K. Vandersypen, and H. Yang.
This research was sponsored  by the Army Research Office (ARO) under Awards No.\ W911NF-17-1-0274 and No.\ W911NF-22-1-0090. 
The views, conclusions, and recommendations contained in this document are those of the authors and are not necessarily endorsed nor should they be interpreted as representing the official policies, either expressed or implied, of the Army Research Office (ARO) or the U.S. Government. The U.S. Government is authorized to reproduce and distribute reprints for Government purposes notwithstanding any copyright notation herein.
This research was performed using the computer resources and assistance of the UW-Madison Center for High Throughput Computing (CHTC) in the Department of Computer Sciences.  The CHTC is supported by UW-Madison, the Advanced Computing Initiative, the Wisconsin Alumni Research Foundation, the Institutes for Discovery, and the National Science Foundation, and is an active member of the Open Science Grid, which is supported by the National Science Foundation and the U.S. Department of Energy's Office of Science.
\end{acknowledgements}

\appendix
 
\section{Magnitude of frequency shift in quantum dot architecture}
\label{magnitude}

Here we give an order-of-magnitude estimate of the PIRS effect and some of the intermediate quantities involved in it for a Si/SiGe hetrostructure quantum dot device with a micromagnet.  
In this case, $\Delta f$ is due to the shift in position of the electron in the non-uniform magnetic field caused by the micromagnet. Let there be an electron spin qubit in a quantum dot at the origin of coordinates.  The electron is at the bottom of a circularly symmetric two-dimensional harmonic potential $k (x^2 + y^2) /2$.  At this point there is a magnetic field gradient $\partial B_i/\partial x_j$, where $i$ and $j$ are Cartesian indices.

The qubit frequency shift is given by $\Delta f = \mathbf{c}_q \cdot [\langle \mathbf{F}_q(T) \rangle - \langle \mathbf{F}_q(T) \rangle (T=0)]$, with \cite{choi2022anisotropy}:
\begin{equation}
    \mathbf{c}_q = \frac{g \mu_B}{h} \frac{q}{m_t \omega_{orb}^2} \ \left( \frac{\partial B_{y}}{\partial x} \hat{x}+\frac{\partial B_{y}}{\partial y} \hat{y}
    \right)
    .
    \label{eq:freq_shift}
\end{equation}

We have assumed that the applied field is in the $y$-direction.  A typical device of this kind that was particularly well-characterized was described in Ref.~\cite{Kawakami:2014p666}. The magnetic field gradients for that device in units of mT(nm)$^{-1}$ were $\partial B_{y}/\partial x = -0.05$ and $\partial B_{y}/\partial y = 0.18$.  The transverse effective mass $m_t = 0.19 \, m_e= 1.73 \times 10^{-31}\,$kg.  We take the lowest orbital excitation frequency as $\omega_{orb} \sim 2 meV/ \hbar$, which is related to the spring constant by $k = m_t \omega_{orb}^2$, and an average value of $|\nabla B|$ as 0.1 mT (nm)$^{-1}$.  These numerical values should be more or less typical of micromagnet-based Si/SiGe devices, but variations from device to device can certainly alter our estimate.

A single component of $ \langle \mathbf{F}_q \rangle$
at $T=0$ is the result of a random-walk summation of the same component of the field exerted at the position of the qubit by the N TLSs.  It is therefore given by $\sqrt{N}$ times the rms value of the individual contributions to one component in the sum in Eq.~1.  There is an additional angular average over the directions of $\mathbf{p}_i$ with the result that $\Delta f(T=0) \sim \sqrt{2N/3}\, p_0 / 4 \pi \varepsilon \varepsilon_r d^3$, where $d$ is an average distance from the TLS to the qubit, and we use $ d \sim 50 $ nm and $\varepsilon_r \sim $11. Combining this with Eq.~\ref{eq:freq_shift}
we find $ \Delta f(T=0) \sim 1.4$ MHz, not too far from what is observed for the maximum $\Delta f$, which should be roughly comarable with the computed quantity.  With these parameters, the qubit moves about 0.5 nm due to $\langle \mathbf{F}_q \rangle$, a field of about 4500 V/m.

\section{Further examples of the temperature dependence of the electric field}
\label{examples}

In this section we give a few representative examples of the temperature dependence of the qubit frequency for single TLS configurations in the different pictures of TLS positions and directions.  In each figure we show two configurations but with the interaction turned off and on.
In fact we compute the $y$ component of $\langle \mathbf{F}_q \rangle$ and leave $\mathbf{c}_q,y$ arbitrary.
Then we enforce the condition that $\Delta f(T=0) = 0 $.  This means that the ground state configurations for the interacting and non-interacting cases and their resultant $\langle \mathbf{F}_q \rangle$ may be quite different.  The curves for the interacting case are smoothed by averaging over the 11 points centered at the plotted point, except at the ends of the curve.  Note that 
$\langle \mathbf{F}_q \rangle \rightarrow 0$ as $T \rightarrow \infty$, but this asymptote is usually off the plotted region.

In Fig.~1 we plot results for the random dipole picture.  Fig.~1(a) is an example where $T^+$ considerably exceeds $T^-$, leading in the non-interacting case to a peak at relatively low T.  Interactions mainly shift the peak but leave non-monotonicity intact.  In Fig.~1(b) the needed cancellation pattern does not occur for the non-interacting case but there is non-monotonicity in the interacting case because of increased cancellation.  Interactions have a strong influence on the ground state configuration for this particular sample, as indicated by the change in sign for the interacting and non-interacting cases. 

In Fig.~2 we see $\Delta f(T)$ for the spherical shell picture for two different TLS configurations.  In this picture the interactions are more effective in producing non-monotonicity and once more we see that the non-monotonicity can be present already in the non-interacting case or it can be induced by the interactions.  $\Delta f(T)$ can show somewhat surprising behavior in the interacting case, in this case a change of sign as a function of T.  This has not been observed to date.  If it were, it would be a sign that interactions are important.

In Fig.~3 we plot $\Delta f(T)$ for two samples belonging to the trap and random dipole pictures.  These plots are mainly included to dispel any impression that non-monotonicity is universal in the IRGM.  Both the non-interacting and the interacting cases can show monotonic behavior as a function of temperature.  This can happen as in Fig.~3(a), where the ground state configuration of the TLS changes drastically when interactions are turned on, or as in Fig.~3(b), where the two ground states are apparently rather similar.    
\begin{figure*}
\centering
\includegraphics[width=\textwidth]{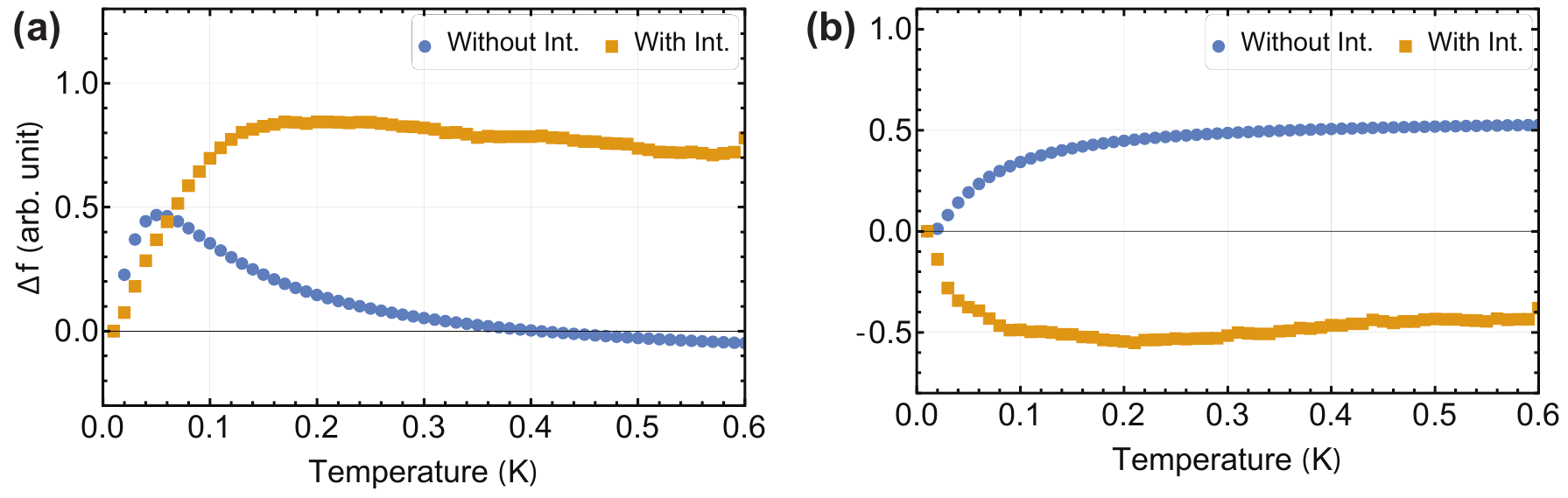}
\caption{Temperature dependence of qubit frequency shifts in the random dipole picture of the TLS.  The shifts are computed from the equilibrium electric field $\langle \mathbf{F}_q \rangle (T)-\langle \mathbf{F}_q \rangle (T=0)$ at the position of the qubit for two configurations of the TLS.  Without Int. and With Int. mean the non-interacting ($H_{int} = 0$) and fully interacting case, respectively.  \textbf{(a)} Both non-interacting and interacting cases show non-monotonic shifts. \textbf{(b)} Example in which the interaction causes non-monotonic behavior.  The applied magnetic field is assumed to be in the $y$-direction.}
     \label{fig:random}
\end{figure*}

\begin{figure*}
\centering
\includegraphics[width=\textwidth]{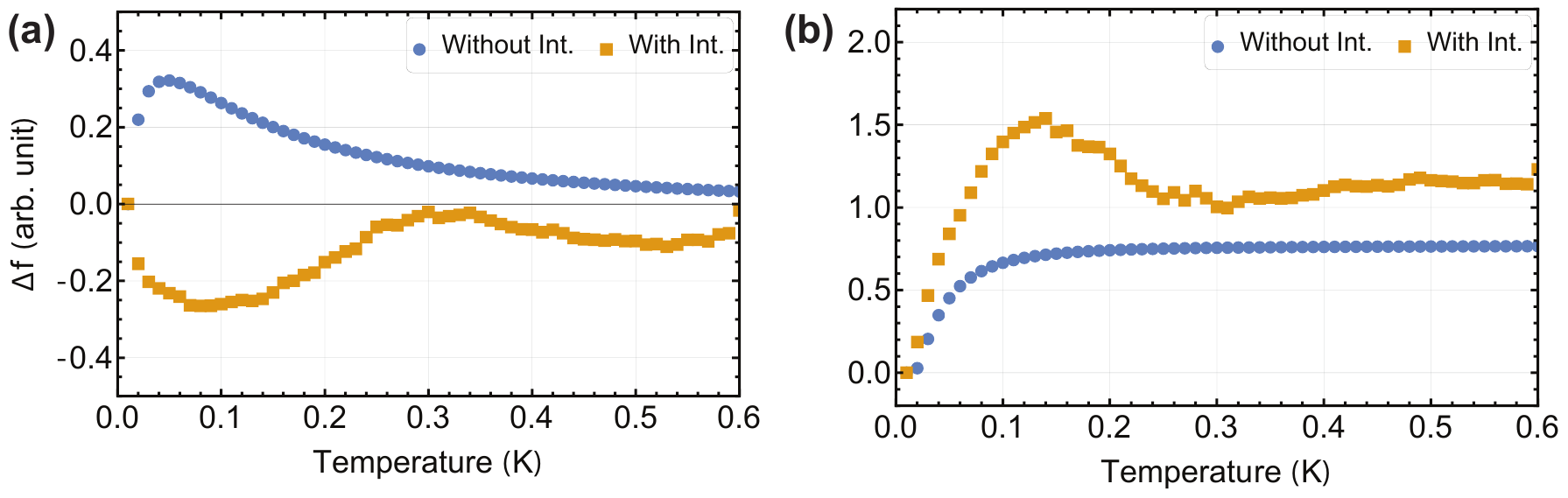}
\caption{Temperature dependence of qubit frequency shifts in the spherical shell picture of the TLS.  The shifts are computed from the equilibrium electric field $\langle \mathbf{F}_q \rangle (T) - \langle \mathbf{F}_q \rangle (T=0)$ at the position of the qubit for two configurations of the TLS.  Without Int. and With Int. mean the non-interacting ($H_{int} = 0$) and fully interacting case, respectively.  \textbf{(a)} Both non-interacting and interacting cases show non-monotonic shifts. \textbf{(b)} Example in which interaction causes non-monotonic behavior. The applied magnetic field is assumed to be in the $y$-direction. Note the change in vertical scale from \textbf{(a)} to \textbf{(b)}.}
     \label{fig:spherical}
\end{figure*}

\begin{figure*}
\centering
\includegraphics[width=\textwidth]{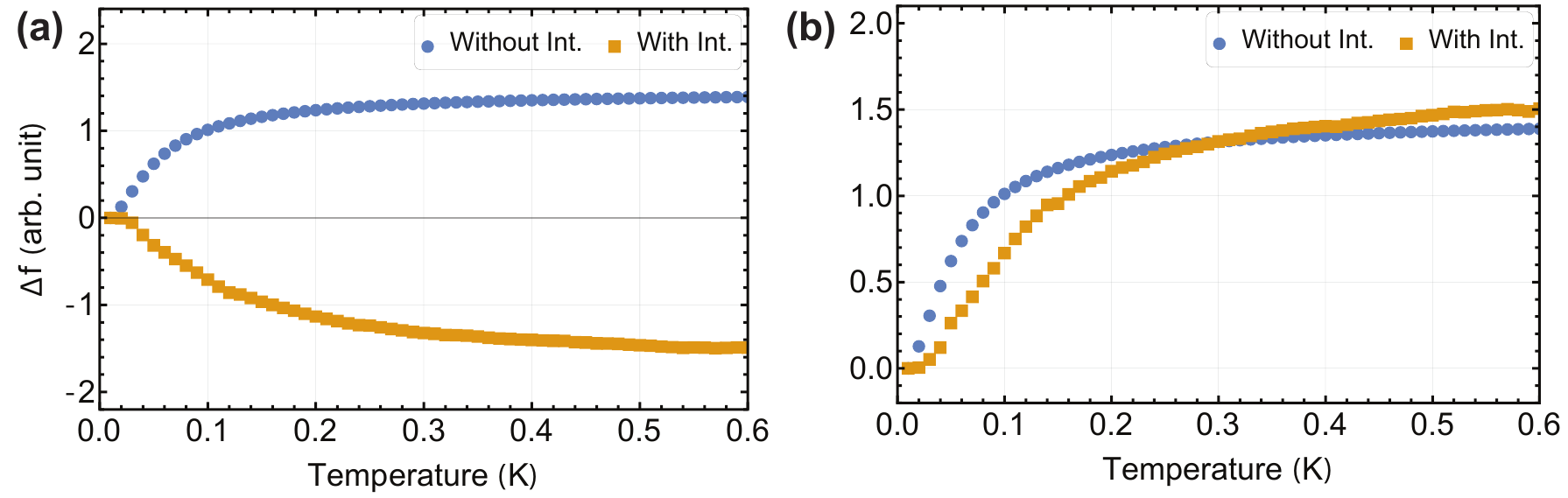}
\caption{Temperature dependence of qubit frequency shifts in the trap and random dipole picture of the TLS.  The shifts are computed from the equilibrium electric field $\langle \mathbf{F}_q \rangle (T) - \langle \mathbf{F}_q \rangle (T=0) $ at the position of the qubit for two configurations of the TLS.  Without Int. and With Int. mean the non-interacting ($H_{int} = 0$) and fully interacting case, respectively.  Both non-interacting and interacting cases show monotonic shifts for \textbf{(a)} trap and \textbf{(b)} random dipole pictures. The applied magnetic field is assumed to be in the $y$-direction. Note the change in vertical scale from \textbf{(a)} to \textbf{(b)}.}
     \label{fig:monotonic}
\end{figure*}

\section{Non-monotonicity criterion}
\label{nonmono}

To determine whether a component $F(T)$ of $\langle \mathbf{F_q} \rangle$  has non-monotonic T dependence, we set up a criterion as follows: from exact or numerical results, we define a set of differences 
\begin{equation}
F_{diff} = \{ F_{m+1} - F_m | \; m = 1,2,3,...,M-1 \}
\end{equation}
where $M = 100$ is the number of temperature points for evaluation.  The average slope magnitude is defined as
\begin{equation}
    s = \frac{1}{M-1} \sum_{m=1}^{M-1} |F_{m+1} - F_m|.
\end{equation}
To avoid false positives from small random fluctuations (particularly important for MC simulations), small slope elements are excluded from $F_{diff}$ so that the smaller set is
\begin{equation}
    F_{large} = \{ F_{m+1} - F_m | \; \frac{s}{2} < |F_{m+1} - F_m| \}.
\end{equation}
Defining signs of the differences as $\sigma_m = \mathrm{sgn} (F_{m+1} - F_m)$, we have groups of positive and negative slopes:
\begin{equation}
\begin{aligned}
F_{pos} &= \{ F_{m+1} - F_m | \; \sigma_m > 0 \; \mathrm{and} \; F_{m+1} - F_m \in F_{large}  \}, \\
F_{neg} &= \{ F_{m+1} - F_m | \; \sigma_m < 0 \; \mathrm{and} \; F_{m+1} - F_m \in F_{large}  \}.
\end{aligned}
\end{equation}
The final non-monotonicity criterion is 
\begin{equation}
    \frac{\min (|F_{pos}|, |F_{neg}|) }{ |F_{pos}|+|F_{neg}| } > 0.1 \; \mathrm{and} \; s > 5
\end{equation}
where the first inequality requires that $\langle \mathbf{F}_q \rangle (T)$ has non-negligible positive and negative parts of slopes and the second one demands that the overall frequency shift in the temperature range is not so small.  The tolerance values, 0.1 and 5, are empirically chosen and can be adjusted for different systems.

\section{Fitting procedure for Fig.~2 of main text}
\label{fitting}

In Fig.~2 of the main text we give a comparison of theory and the PIRS experiment of Undseth \textit{et al.} in which $\Delta f$ was measured for each of six qubits in a row.  We found that the best fit is obtained by taking the case that $T_r \gg T_{int}$, which amounts to a non-interacting model.  We used a trap picture, but the other two pictures could also have been used for the fit.  The parameters are the same as those in the description of  the trap picture in the main text except for the $z$ coordinates of the TLS, which are now taken as $z_j = 36$ nm following Ref.~\cite{Undseth, philips2022universal}, and $\Delta E_0 = 10^5$ V/m.  The six samples are generated by varying the conversion factor $\mathbf{c}_{q,y}$ and the positions of TLS in the $x-y$ plane, which are randomly assigned within circles whose centers are TLS positions of a reference sample and radii are 5 nm.  The applied magnetic field is assumed to be in the $y$-direction.  As we showed in Sec.~1, the order of magnitude of the effect is consistent between theory and experiment.  The best fit for the conversion factors for the qubits $Q_i$ with $i =$ 1, 2, 3, 4, 5, 6 were in the ratios 0.105 : 0.114 : 0.072 : 0.077 : 0.103 : 0.059. All fit parameters are available from the authors on request.

\vspace{20mm}
\bibliography{main}

\end{document}